\documentstyle[preprint,eqsecnum,aps]{revtex}

\def\sqr#1#2{{\vcenter{\hrule height.#2pt\hbox{\vrule width.#2pt
height#1pt \kern#1pt \vrule width.#2pt}\hrule height.#2pt}}}

\begin{document}
\draft
\title{Static black holes of metric--affine gravity in the presence of matter}
\author{Eloy Ay\'on--Beato$^{\P}$\thanks{E--mail: ayon@fis.cinvestav.mx},
Alberto Garc\'{\i}a$^{\P}$\thanks{E--mail: aagarcia@fis.cinvestav.mx},
Alfredo Mac\'{\i}as$^{\diamond}$\thanks{E--mail: amac@xanum.uam.mx} and
Hernando Quevedo$^\star$\thanks{E--mail: quevedo@nuclecu.unam.mx}}
\address{$^\P$~Departamento~de~F\'{\i}sica,~Centro~de~%
Investigaci\'on~y~de~Estudios~Avanzados~del~IPN,\\
Apartado~Postal~14--740,~C.P.~07000,~M\'{e}xico,~D.F.,~MEXICO.\\
$^{\diamond}$~Departamento~de~F\'{\i}sica,~Universidad~Aut%
\'onoma~Metropolitana--Iztapalapa\\
Apartado~Postal~55--534,~C.P.~09340,~M\'{e}xico,~D.F.,~MEXICO.\\
$^\star$~Instituto~de~Ciencias~Nucleares,~Universidad~Nacional~Aut%
\'onoma~de~M\'exico\\
Apartado~Postal~70--543,~C.P.~04510,~M\'{e}xico,~D.F.,~MEXICO.}

\date{\today}
\maketitle

\begin{abstract}
We investigate spherically symmetric and static gravitational
fields representing black hole configurations in the framework of
metric--affine gauge theories of gravity (MAG) in the presence of
different matter fields. It is shown that in the {\it triplet
ansatz} sector of MAG, black hole configurations in the presence
of non--Abelian matter fields allow the existence of black hole
hair. We analyze several cases of matter fields characterized by
the presence of hair and for all of them we show the validity of
the no short hair conjecture.\\ {\bf file nh3.tex, 16.04.2001}
\end{abstract}

\pacs{PACS numbers: 0450, 0420J, 0350K}


\section{Introduction}

One of the first attempts to consider a non--Riemannian
description of the gravitational field coupled to an
electromagnetic field is due to Weyl \cite{Weyl}. Although this
approach is considered today as unsuccessful, recent developments
through a theory that unifies all the fundamental interactions are
reviving the interest in non--Riemannian structures. For instance,
the unification scheme in the framework of string theory indicates
that the classical Riemannian description is not valid on all
scales.

Indeed, the theory of the quantum superstring \cite{quantum}
indicates that non--Riemannian features are present on the scale
of the Planck length. It turns out that low--energy dilaton and
axi--dilaton interactions are tractable in terms of a connection
that leads to a non--Riemannian geometrical structure with a
particular torsion and nonmetricity fields. Therefore, it is
interesting to investigate gravity theories which generalize the
pure Riemannian geometrical structure of Einstein's theory.

MAG is a gauge theory of the 4--dimensional affine group endowed
with a metric. As a gauge theory, it finds its appropriate form if
expressed with respect to arbitrary frames or {\em coframes}. The
corresponding gravitational potentials are the metric
$g_{\alpha\beta}$, the coframe $\vartheta^\alpha$, and the
connection 1--form $\Gamma_\alpha{}^\beta$, with values in the Lie
algebra of the 4--dimensional linear group $GL(4,R)$. Therefore,
spacetime is described by a metric--affine geometry with the
gravitational {\em field strengths} nonmetricity
$Q_{\alpha\beta}:=-Dg_{\alpha\beta}$, torsion
$T^\alpha:=D\vartheta^\alpha$, and curvature $R_\alpha{}^\beta:=
d\Gamma_\alpha{}^\beta
-\Gamma_\alpha{}^\gamma\wedge\Gamma_\gamma{}^\beta$. Thus, the
post--Riemannian components, nonmetricity and torsion, are
dynamical variables which together with the metric and the
connection provide an alternative description of gravitational
physics.

In this work, we consider MAG as a gravity theory and investigate
static spherically symmetric fields describing black hole
configurations. In particular, we investigate black holes of the
triplet ansatz sector of MAG in the presence of matter represented
by a $SU(2)$ Yang--Mills, Skyrme, Yang--Mills--dilaton,
Yang--Mills--Higgs and a non--Abelian Proca field. We first show
that, in general, the presence of matter with non--Abelian
structure does not allow to apply the arguments used to prove
no--hair theorems. We show that in all these cases black hole hair
exists and it must extend beyond a surface situated at 3/2 the
horizon radius. i.e., we prove the validity of the {\it no short
hair conjecture} in this sector of MAG.

In Section II we review the main aspects of MAG and its {\em
triplet ansatz} sector together with Obukhov's equivalence
theorem. Besides the pure geometric components of MAG, we consider
an additional matter field. In Section III we analyze the general
equations of motion for non--Abelian  matter fields, consider a
static spherically symmetric configuration under the assumption
that it represents the gravitational field of a black hole, and
prove the validity of the {\it no short hair conjecture}. Finally,
in Section IV we discuss our results.


\section{Field equations of MAG and the triplet ansatz}

Let us consider a frame field and a coframe field denoted by
\begin{equation}
e_{\alpha}=e^{\mu}{}_{\alpha}\,\partial _{\mu}\, , \qquad \qquad
\vartheta^{\beta}=e_{\mu}{}^{\beta}dx^{\mu}
\, ,\label{coframe}
\end{equation}
respectively.
The $GL(4,R)$--covariant derivative for a tensor valued $p$--form is
\begin{equation}
D=d+\Gamma_{\alpha}{}^{\beta}\,\,\rho(L^\alpha{}_{\beta})\,\wedge\,,
\label{covariantder}\end{equation} where
$\rho(L^\alpha{}_{\beta})$ is the representation of $GL(4,R)$ and
$L^{\alpha}{}_{\beta}$ are the generators; the connection
one--form is $\Gamma_{\alpha}{}^{\beta}=
\Gamma_{\mu\alpha}{}^{\beta}dx^{\mu}$. The nonmetricity one--form,
the torsion and curvature two--forms read
\begin{equation}
Q_{\alpha\beta}:=-Dg_{\alpha\beta}\, , \quad
T^{\alpha}:=D\vartheta^{\alpha}\, , \quad
R_{\alpha}{}^{\beta}:= d\Gamma_{\alpha}{}^{\beta} -
  \Gamma_{\alpha}{}^{\gamma} \wedge \Gamma_{\gamma}{}^{\beta}\, ,
\label{curv}
\end{equation}
respectively, and the Bianchi identities are
\begin{equation}\label{bianchi}
DQ_{\alpha\beta}\equiv 2R_{(\alpha\beta)}\, , \quad
DT^{\alpha}\equiv R_{\gamma}{}^{\alpha}\wedge\vartheta^{\gamma}\,,
\quad
  DR_{\alpha}{}^{\beta}\equiv 0\, .
\end{equation}
It is worthwhile to stress the fact that
$Q_{\alpha\beta}$, $T^{\alpha}$ and
$R_{\alpha}{}^{\beta}$ play the role of field strengths.

We will consider a metric--affine theory described by
the particular Lagrangian
\begin{equation}\label{Ltot}
{\cal L}=V_{\rm MAG}+ {\cal L}_{\rm MAT}\,,
\end{equation}
where ${\cal L}_{\rm MAT}$ represents the Lagrangian of the matter
field.

In a metric--affine spacetime, the curvature has {\em eleven}
irreducible pieces \cite{PR}, whereas the nonmetricity has {\em
four} and the torsion {\em three} irreducible pieces. The most
general parity conserving Lagrangian $V_{\rm MAG}$ which has been
constructed in terms of all irreducible pieces of the
post--Riemannian components has been investigated previously
\cite{hema99} and reads:

\begin{eqnarray}
\label{QMA} V_{\rm MAG}&=&
\frac{1}{2\kappa}\,\left[-a_0\,R^{\alpha\beta}\wedge\eta_{\alpha\beta}
  -2\lambda\,\eta+T^\alpha\wedge{}^*\!\left(\sum_{I=1}^{3}a_{I}\,^{(I)}
    T_\alpha\right)\right.\nonumber\\ &+&\left.
  2\left(\sum_{I=2}^{4}c_{I}\,^{(I)}Q_{\alpha\beta}\right)
  \wedge\vartheta^\alpha\wedge{}^*\!\, T^\beta + Q_{\alpha\beta}
  \wedge{}^*\!\left(\sum_{I=1}^{4}b_{I}\,^{(I)}Q^{\alpha\beta}\right)\right.
\nonumber \\&+&
b_5\bigg.\left(^{(3)}Q_{\alpha\gamma}\wedge\vartheta^\alpha\right)\wedge
{}^*\!\left(^{(4)}Q^{\beta\gamma}\wedge\vartheta_\beta \right)\bigg]
\nonumber\\&- &\frac{1}{2\rho}\,R^{\alpha\beta} \wedge{}^*\!
\left(\sum_{I=1}^{6}w_{I}\,^{(I)}W_{\alpha\beta}
  +w_7\,\vartheta_\alpha\wedge(e_\gamma\rfloor
  ^{(5)}W^\gamma{}_{\beta} ) \nonumber\right.\\&+& \left.
  \sum_{I=1}^{5}{z}_{I}\,^{(I)}Z_{\alpha\beta}+z_6\,\vartheta_\gamma\wedge
  (e_\alpha\rfloor ^{(2)}Z^\gamma{}_{\beta}
  )+\sum_{I=7}^{9}z_I\,\vartheta_\alpha\wedge(e_\gamma\rfloor
  ^{(I-4)}Z^\gamma{}_{\beta} )\right)
\label{6}\,.
\end{eqnarray}
The Minkowski metric is $o_{\alpha\beta} = \hbox{diag}(-+++)$,
$^*$ is the Hodge dual, $\eta:={}^*\!\, 1$ is the volume
four--form, the constant $\lambda$ is the cosmological constant,
$\rho$ the strong gravity coupling constant, the constants $ a_0,
\ldots a_3$, $b_1, \ldots b_5$, $c_2, c_3,c_4$, $w_1, \ldots w_7$,
$z_1, \ldots z_9$ are dimensionless. We have introduced in the
curvature square term the irreducible pieces of the antisymmetric
part $W_{\alpha\beta}:= R_{[\alpha\beta]}$ and the symmetric part
$Z_{\alpha\beta}:= R_{(\alpha\beta)}$ of the curvature 2--form. In
$Z_{\alpha\beta}$, we have the purely {\em post}--Riemannian part
of the curvature. Note the peculiar cross terms with $c_I$ and
$b_5$.

We will consider here only
the simplest non--trivial case of torsion and nonmetricity
with shear. Then, for the nonmetricity we use the ansatz
\begin{equation}
  Q_{\alpha\beta}=\, ^{(3)}Q_{\alpha\beta} +\,
  ^{(4)}Q_{\alpha\beta}\, ,\label{QQ}
\end{equation}
where
\begin{equation}
  ^{(3)}Q_{\alpha\beta}={4\over
    9}\left(\vartheta_{(\alpha}e_{\beta)}\rfloor \Lambda - {1\over
      4}g_{\alpha\beta}\Lambda\right)\,,\qquad \hbox{with}\qquad
  \Lambda:= \vartheta^{\alpha}e^{\beta}\rfloor\!
  {\nearrow\!\!\!\!\!\!\!Q}_{\alpha\beta}\label{3q}\,,
\end{equation}
is the proper shear piece and $^{(4)}Q_{\alpha\beta}=
Q\,g_{\alpha\beta}$ represents the dilation piece, where
$Q:=(1/4)\,Q_{\gamma}{}^{\gamma}$ is the Weyl one--form, and
${\nearrow\!\!\!\!\!\!\!Q}_{\alpha\beta}
:=Q_{\alpha\beta}-Q\,g_{\alpha\beta}$ is the traceless piece of
the nonmetricity. Other pieces of the irreducible decomposition of
the nonmetricity \cite{PR} are taken to be zero.

Let us choose for the torsion only
the covector piece as non--vanishing:
\begin{equation}
T^{\alpha}={}^{(2)}T^{\alpha}={1\over 3}\,\vartheta^{\alpha}\wedge
T\,,
\qquad \hbox{with}\qquad T:=e_{\alpha}\rfloor T^{\alpha}\,.\label{TT}
\end{equation}
Thus we are left with a triplet of non--trivial one--forms $Q$,
$\Lambda$, and $T$ for which we make the following ansatz
\begin{equation}
Q=\frac{k_0}{k_1}\Lambda
=\frac{k_0}{k_2}T
\label{genEug}\, ,
\end{equation}
where $k_0$, $k_1$ and $k_2$ are given in terms of the
gravitational coupling constants (for details, see \cite{hema99}).
This is the so--called {\em triplet ansatz} sector of MAG theories
\cite{obu,tw,ghlms}.

Consequently, here, we limit ourselves to the special case in
which the only surviving strong gravity piece is the square of the
segmental curvature (with vanishing cosmological constant), i.e.
\begin{eqnarray}
V_{MAG}&=& \frac{1}{2\kappa}\,\left[-a_0\,R^{\alpha\beta}\wedge
\eta_{\alpha\beta}+
a_{2}\,T^\alpha\wedge{}^*\!\,^{(2)}T_\alpha\right.
\nonumber\\&+&\left.
2\left(c_{3}\,^{(3)}Q_{\alpha\beta}+c_{4}\,^{(4)}
Q_{\alpha\beta}\right)\wedge\vartheta^\alpha
\wedge{}^*\! T^\beta\right.\nonumber\\& +&
\left.Q_{\alpha\beta}\wedge{}^*\!\left(b_{3}\,^{(3)}
Q^{\alpha\beta}+b_{4}\,^{(4)}Q^{\alpha\beta}\right)\right]
\nonumber\\&-& \frac{z_{4}}{2\rho}\,R^{\alpha\beta}
\wedge{}^*\!\,^{(4)}Z_{\alpha\beta}
\label{nondeg1}\,,
\end{eqnarray}
where
\begin{equation}
-\frac{z_4}{2\rho}\, R_\alpha{}^\alpha\wedge\,^\ast Z_\beta{}^\beta=
  -\frac{2 z_4}{\rho}\,dQ\wedge{}^\ast dQ\,\label{seg^2}
\end{equation}
is the kinetic term for the Weyl one--form.

Under the above given assumptions it is now straightforward to
apply Obukhov's equivalence theorem \cite{obu,tw,mms98} according
to which the field equations following from the pure geometrical
part of the Lagrangian (\ref{Ltot}), i.e., $V_{MAG}$, are
equivalent to Einstein's equations with an energy--momentum tensor
determined by a Proca field. In the case investigated here we have
an additional term due to the presence of the matter field in
(\ref{Ltot}). Thus, the field equations read
\begin{eqnarray}\label{fieldob}
\frac{a_0}{2}\,\eta_{\alpha\beta\gamma}\wedge\tilde{R}^{\beta\gamma}
&=& \kappa\,\Sigma_\alpha\,,\\
d\,{}^{\ast }\!H+m^{2}\,{}^{\ast }\!\phi &=&0,  \label{eq:Proca0}
\end{eqnarray}
where $\phi$ represents the Proca 1--form, $H\equiv d\phi $, $m$
is completely given in terms of the coupling constants, and a
tilde denotes the Riemannian part of the curvature.
The energy--momentum current entering the right hand
side of the Einstein equations is given by
\begin{equation}
\Sigma_\alpha = \Sigma_\alpha^{(\phi)} +
\Sigma_\alpha^{(\rm MAT)} \ ,
\end{equation}
where
\begin{eqnarray}
  \Sigma_\alpha^{(\phi)}& := \frac{z_4k_0^2}{2\rho} \left\{ \left(
      e_\alpha\rfloor d\phi \right)\wedge{}^\ast d\phi- \left(
      e_\alpha\rfloor\,^\ast d\phi \right)\wedge\, d\phi \right.
  \nonumber \\ &\quad\left. +\;m^2\,\left[ (e_\alpha\rfloor
      \phi)\wedge{}^\ast \phi\;+\;
      (e_\alpha\rfloor\,^\ast\phi)\wedge{}\phi\right] \right\}\,
               \label{ProcaEM}
\end{eqnarray}
is the energy--momentum current of the Proca field, and
$\Sigma_\alpha^{(\rm MAT)}$ is energy--momentum current of the
additional matter field which satisfies also the corresponding
Euler--Lagrange equations.

Thus, the triplet ansatz sector of a MAG theory
coupled to a matter field has been reduced
to the effective Einstein--Proca system of differential equations
coupled to a matter field.


\section{Static black holes in MAG}

In a recent work \cite{eaah}, we have investigated the
gravitational field configuration corresponding to static
spherically symmetric black holes in the context of the triplet
ansatz sector of MAG, and we have proven a no-hair theorem for
this specific case. It was shown that for the case of a massive
Proca field ($m\neq 0$) in the presence of a static black hole,
the effective Proca field is trivial and the field equations
reduce to the vacuum Einstein equations and, hence, the only
static black hole is described by the Schwarzschild solution.
Moreover, for a massless Proca field ($m=0)$, the equations reduce
to the Einstein--Maxwell system and, therefore, the
Reissner--Nordstr\"om solution is the only static black hole with
non--degenerate horizon.

In addition, we have pointed out that for spherically symmetric
static configurations of the triplet ansatz sector of MAG coupled
to a Maxwell field, the only black hole solution allowed is the
Reissner--Nordstr\"om one, because in this case the field
equations are equivalent to an effective Einstein--Proca--Maxwell
system. The question arises: Are these no--hair theorems valid
also when the geometrical components of the triplet ansatz sector
of MAG become (minimally) coupled to a different kind of matter
fields?

\subsection{Matter fields with non--Abelian structure}

 In this section, we will show that our no--hair theorems are
not valid in the presence of matter fields characterized by
a non--Abelian gauge structure.

Most of the proofs of no--hair theorems are based upon the method
first developed by Bekenstein \cite{bek} which consists on
rearranging the field equations into a statement about the
behavior of fields outside the event horizon. We have improved
this method in our previous work \cite{eaah} to prove the no--hair
theorems for the triplet ansatz sector of MAG. Following
\cite{GreeneMO'N93} we will show that when an additional
non--Abelian matter field is taken into account, the original
argument for no--hair can be avoided.

Consider a set of arbitrary matter fields $\Psi_i$ in a
gravitational background described by the effective
Einstein--Proca system of MAG. The corresponding action becomes $S
= \int (V_{\rm MAG} + {\cal L}_{\rm MAT}) d^4 x$. After
multiplication by $\Psi_i d^4 x$ and integration, the
Euler--Lagrange equations for the matter fields can be written as
\begin{equation}
\sum_i \int_\Omega \partial_\mu
\left[ \Psi_i {\partial {\cal L}_{\rm MAT}\over
 \partial (\partial_\mu \Psi_i)}\right] d^4 x =
\sum_i \int_\Omega \left[
\Psi_i {\partial {\cal L}_{\rm MAT} \over \partial \Psi_i }
+
\partial_\mu(\Psi_i ){\partial {\cal L}_{\rm MAT}
\over \partial(\partial_\mu \Psi_i )}\right] d^4 x \ .
\label{eula}
\end{equation}
Assuming that this coupled system admits black hole solutions,
then the left--hand side of Eq.~(\ref{eula}) can be expressed as a
surface integral over the hypersurface $\partial\Omega$ which
bounds the volume $\Omega$ exterior to the black hole. As has been
shown by Bekenstein \cite{bek}, this surface integral vanishes for
static fields if the ``norm" of the integrand
\begin{equation}
\sum_{i,j} g_{\mu\nu}\Psi_i \Psi_j
{\partial{\cal L}_{\rm MAT}\over \partial (\partial_\mu \Psi_i) }
{\partial{\cal L}_{\rm MAT}\over \partial (\partial_\nu \Psi_j) }
\end{equation}
is finite on the horizon. Furthermore, if one can show that the
integrand of the right hand side of Eq.~(\ref{eula}) is either
positive or negative definite, then the only solutions with finite
enegy are those for which the integrand vanishes. This is the most
used method to prove no--hair theorems in different theories (see
Ref.~\cite{Ayon00} for a more recent approach).

Consider the Lagrangian for an Abelian Yang--Mills matter field:
$ {\cal L}_{\rm MAT} =\sqrt{-g}|F|^2 =
 \sqrt{-g}
F^{\ \ a}_{\mu\nu}F^{\mu\nu}_{\ \ a}$ where $F^{\ \ a}_{\mu\nu}=
2\partial_{[\mu} A^{\ \ a}_{\nu]}$ and $a$ is the internal index.
(The coupling constants are irrelevant for our analysis.) The
calculation of the right--hand side of Eq.~(\ref{eula}) yields
$2\int_\Omega \sqrt{-g} |F|^2d^4 x$. If we consider a static field
and, for the sake of simplicity without loss of generality, assume
that the time component $A^{\ a}_t =0$ , then $|F|^2 \geq 0$. This
shows that the argument for no--hair can be applied in this case.

Consider now the Lagrangian for a non--Abelian Yang--Mills matter
field. Due to the nonlinear terms of the field strength ($e$ is
the gauge coupling constant)
\begin{equation}
 F^{\ \ a}_{\mu\nu}= 2\partial_{[\mu} A^{\ a}_{\nu]}
+ e \epsilon^a_{\ bc} A^{\ b}_\mu A^{\ c}_\nu  \ ,
\end{equation}
the right--hand side of Eq.~(\ref{eula}) gives
\begin{equation}
2 \sum_a \int_\Omega \sqrt{-g} \left[  |F|^2
+ e \epsilon_{abc} A^{\ b}_\mu A^{\ c}_\nu F^{\mu\nu\, a}
\right] \ .
\label{eulana}
\end{equation}
As in the Abelian case, one can show that $|F|^2 \geq 0$ for
static fields. However, the second term in the integrand of
Eq.~(\ref{eulana}) has not a definite sign, but depends on the
particular solution for the exterior field of the black hole. This
opens the possibility of avoiding the original no--hair argument.
That is, the non--Abelian structure of the matter field can affect
the statement about no--hair in the exterior of a static black
hole. It is worth  mentioning that a similar argument can be used
to show that an additional potential term in the matter Lagrangian
of an Abelian Yang--Mills field can affect the no--hair statement
in the same way as does the non--Abelian structure of the field by
itself.

\subsection{Black holes with hair}

In all theories in which black hole solutions with hair have been
discovered, only the special case of static spherically symmetric
spacetimes has been analyzed. For this reason, we will now
investigate the triplet ansatz sector of MAG for this specific
case. In the standard tensor notation, the field equations for the
effective Einstein--Proca sector of MAG (\ref{fieldob}) can be
written as follows
\begin{equation}
R_{\mu \nu } -{1\over 2}g_{\mu\nu} R =
{ \widetilde{\kappa } }
\left(\Sigma_{\mu\nu}^{(\phi)}+\Sigma_{\mu\nu}^{\rm MAT}\right) \ ,
\label{eq:Ein}
\end{equation}
with
\begin{equation}
\Sigma_{\mu\nu}^{(\phi)}=
  H_{\mu }^{~\lambda }H_{\nu \lambda }
- \frac{1}{4}g_{\mu \nu }H_{\lambda\tau }H^{\lambda \tau }
+ m^{2}\phi _{\mu }\phi _{\nu }
- {m^2\over 2}g_{\mu\nu}\phi_\lambda \phi^\lambda
\, ,  \label{emproca}
\end{equation}
where $R_{\mu \nu }$ is the Ricci tensor, $R$ is the curvature
scalar, $H_{\mu \nu }\equiv 2\nabla _{\lbrack \mu }\phi _{\nu ]}$
is the field strength of the {\it Abelian} Proca field
$\phi_{\mu}$, and
$\widetilde{\kappa}\equiv\kappa{z}_4k_0^2/4\pi\rho{a}_0$.
Moreover, the Proca field must satisfy the motion equation
(\ref{eq:Proca0}) which in tensor notation reads
\begin{equation}
\nabla _{\nu }H^{\nu \mu }=m^{2}\phi ^{\mu }\ .
\label{eq:Proca}
\end{equation}
Finally, the energy--stress tensor for the additional matter
$\Sigma^{\rm MAT}_{\mu\nu}$ can be explicitly calculated from
the corresponding matter Lagrangian.

We will consider
asymptotically flat static spherically symmetric black hole
spacetimes and write the corresponding coframe as
\begin{equation}
\vartheta ^{\hat{0}} =\, e^{-\delta}\, \mu^{1/2}d\, t \, ,\quad
\vartheta ^{\hat{1}} =\, \mu^{-1/2}\, d\, r\, , \quad
\vartheta ^{\hat{2}} =\,  r\, d\,\theta\,,\quad
\vartheta ^{\hat{3}} =\,  r\, \sin\theta \, d\,\varphi
  \,,\label{frame2}
\end{equation}
where $\delta$ and  $\mu=1-2M(r)/r$ are functions of $r$ only. The
coframe is assumed to be {\em orthonormal} with the local
Minkowski metric $o_{\alpha\beta}:=\hbox{diag}(-1,1,1,1)
=o^{\alpha\beta}$. The condition that the metric corresponding to
the coframe (\ref{frame2}) describes the gravitational field of a
black hole implies that there exists  a regular event horizon at a
finite distance, say $r_H$, so $M(r_H)=r_H/2$, and $\delta(r_H)$
must be finite. On the other hand, asymptotic flatness requires
that $\mu \to 1$ and $\delta \to 0$, at infinity. We also assume
that the Proca field as well as the matter fields to be considered
below respect the symmetries of the spacetime, i.e. they are
static and spherically symmetric.

The Einstein's equations together with the equations of motion for
the Proca and matter fields (\ref{fieldob}) form a dependent set
as they are related by the Bianchi identities which in this case
can be written in the form of a conservation law $\nabla_\mu
\Sigma^{\mu\nu}=0$. This conservation equation has only one
non--trivial component (the $r$ component) which can be written
as:
\begin{equation}
e^{\delta} (e^{-\delta} \Sigma^{r}_{r})^{'} = {1 \over 2\mu r}
[(\Sigma^{t}_{t} - \Sigma^{r}_{r})
+ \mu(2\Sigma -3(\Sigma^{t}_{t} +\Sigma^{r}_{r}))] \ .
\label{eq:con}
\end{equation}
Moreover, the field equations (\ref{eq:Ein}) for the coframe
(\ref{frame2}) yield
\begin{eqnarray}
\mu^\prime&=&\widetilde{\kappa }
\,r\,{\Sigma^t}_t + {{1-\mu}\over r},\label{eq:mu}\\
\delta^\prime&=&{{\widetilde{\kappa }
  \,r}\over{2\,\mu}}\,({\Sigma^t}_t-{\Sigma^r}_r ),\label{eq:delta}
\end{eqnarray}
where the prime stands for differentiation with respect to the
radial coordinate $r$, and $\Sigma_{\mu\nu} = \Sigma_{\mu\nu}^{(\phi)}
+ \Sigma_{\mu\nu}^{\rm MAT}$.

The set of equations (\ref{eq:con})--(\ref{eq:delta}) have been
intensively analyzed in the literature for different theories and
numerical solutions have been found that are characterized by the
presence of non--Abelian and Higgs hair. These theories are:
$SU(2)$ Yang--Mills, Skyrme, Yang--Mills--dilaton,
Yang--Mills--Higgs and non--Abelian Proca. Because of the
additivity of the energy--momentum tensor, hair will also exist in
any theory which involves an arbitrary combination of the matter
fields mentioned above. This is true also for any linear
combination of energy--stress tensors in which at least one of
them is characterized by the presence of hair. Accordingly, black
hole hair will exist when any one of these matter fields is
present in the gravitational field of a static spherically
symmetric black hole described by the triplet ansatz sector of
MAG.


\subsection{The no short hair conjecture in MAG}

In a recent work \cite{ddh}, it was conjectured that {\it if a
black hole has hair, then it cannot be shorter than the
radius $r_{\rm hair} = 3/2\sqrt{A/4\pi}$, where $A$ is the
horizon area}. In the case of a static spherically symmetric
black hole the hair radius $r_{\rm hair} = 3/2 r_{\rm H}$, where
$r_{\rm H}$ is the horizon radius. The hair radius defines
around a black hole a hypersurface, called the ``hairosphere",
beyond which the hair can exist. Inside the hairosphere, hair
is not allowed to exist. In this section we will show that
this {\it no short hair conjecture} is valid for the special
case of MAG under consideration in presence of the matter
fields in which black hole hair has been discovered.

In all cases we will consider, the effective Einstein--Proca
field equations of MAG (\ref{eq:mu}) and (\ref{eq:delta})
and the corresponding ansatz for the matter
fields can be written as:
\begin{equation}
\mu^\prime    =  {1\over r}[ 1-\mu  +\alpha \,( K  + U)],\qquad
\delta^\prime  =  \beta\, K \ ,
\label{eq:ee}
\end{equation}
where $\alpha, \beta,\,  K $ and $U$ take particular values in
each case. The strategy for showing the validity of the no short
hair conjecture is the following. From the matter field equations
and Eq.~(\ref{eq:ee}) it is possible to obtain a generic function
\cite{Su} of the form $E \propto e^{-\delta} ( K  - U) $ which
enters the conservation equation (\ref{eq:con}). If we demand the
existence of a black hole solution in each case, we will show that
$E$ must be negative on the horizon and positive semidefinite at
infinity. Therefore, $E$ must be positive in some region (between
the horizon and infinity) which is determined by the condition
$3\mu > 1$. Finally, we show that this region corresponds to
values of the radial coordinate $r$ outside the hairosphere. This
proves the validity of the conjecture.

We now investigate all the particular cases in which black hole
hair has been found. For the sake of simplicity, in each case we
will quote for $\alpha,\ \beta, \ K$ and $U$ in Eq.~(\ref{eq:ee})
only the term corresponding to the additional matter field,
dropping the term coming from the effective Proca field of MAG
which does not allow the presence of hair \cite{eaah}. Because of
the additivity of the stress--energy tensor discussed above, this
simplification does not affect the behavior of the generic
function $E$.

$i$) The $SU(2)$ Yang-Mills \cite{ym} field for which the matter
Lagrangian has the form
\begin{equation}
{\cal L}_{\rm MAT} ={\cal L}_{\rm YM} =
-{\sqrt{-g}\over 16\pi f^2}
 {F_{\mu\,\nu}}^a\,{F^{\mu\,\nu}}_a,
\end{equation}
where ${F_{\mu\,\nu}}^a=\partial_\mu {A_\nu}^a - \partial_\nu
{A_\mu}^a + {\epsilon^a}_{bc}{A_\mu}^b\,{A_\nu}^c$ is the field
strength for the gauge field ${A_\mu}^a$, and $f$ represents the
gauge coupling constant. We use the static spherically symmetric
ansatz for the potential
\begin{equation}
A=\sigma_a\,{A_\mu}^a\,dx^\mu=\sigma_1 \,w\,d\,\theta
+ (\sigma_3\,\cot\theta +
\sigma_2\,w)\sin\theta\,d\varphi,
\label{eq:pot}
\end{equation}
where $\sigma_i\ (i=1,2,3)$ are the Pauli matrices and $w$ is a
function of $r$ only. The field equations for this case may be
written as in Eq.~(\ref{eq:ee}), with $ K  = \mu \,{w'}^2$, $U =
(1-w^2)^2/ (2r^2)$, $\alpha=  -2/f^2$, and $  \beta =-2/(f^2 \mu
r)$. From the matter field equations we obtain
\begin{equation}
E' \equiv [ r^2 e^{-\delta} (K - U) ]'
= r e^{-\delta} (3\mu - 1) w'^2 \ . \label{eq:eym}
\end{equation}

From the expressions for $ K $ and $ U$ we see that $ E$ is
negative at the horizon because  $ K (r_H) =0$, (since $\mu(r_H) =
0$),  and $U(r_H) > 0$. On the other hand, the asymptotic flatness
condition implies that  $ E \rightarrow 0$ as $ r\rightarrow
\infty$. Accordingly, $ E$ must be an increasing function of $r$
in some intermediate region. It follows then that the right hand
side of Eq.~(\ref{eq:eym}) must become  positive at some point,
i.e. we must have  $3\mu >1$.

$ii$) The Skyrme field with the matter Lagrangian  \cite{es}
\begin{equation}
{\cal L}_{\rm MAT}={\cal L}_{\rm SK}=
\sqrt{-g} {f^2\over 4}\,
Tr( \nabla_\mu W \nabla^\mu W^{-1}) +
{\sqrt{-g} \over 32\,e^2}
Tr[ (\nabla_\mu W)\,W^{-1}, (\nabla_\nu )\,W^{-1}]^2,
\end{equation}
where $\nabla_\mu$ is the covariant derivative, $W$ is the $SU(2)$
chiral field, and $f^2$ and $ e^2$ are the coupling constants.
For the $SU(2)$ chiral field we use the {\em hedgehog} ansatz
$ W(r) = exp({\bf \sigma \cdot r}\, F(r)) $ where ${\bf \sigma}$ are
the Pauli matrices and ${\bf r}$ is a unit radial vector.

To write down the field equations we follow \cite{es} and use the
variables $\tilde r = e\,f\,r$, and $\tilde m (\tilde r) = efm(r)$
so that the function $\mu$ defined above remains invariant.
Dropping the tilde, the resulting equations are equivalent to
Eq.~(\ref{eq:ee}) with $ K = \mu\,[r^2/2 + \sin^2 F(r)] F'^2$, $U
= \sin^2 F\,[1+\sin^2F\, /(2\,r^2)]$, $\alpha = -8\pi f^2$, and
$\beta =-8\pi f^2/(\mu r)$. From the matter field equations we
find
\begin{equation}
 E' \equiv  [ e^{-\delta} ( K  - U)]' =
- e^{-\delta} \left[ r\mu F'^2 + { 1-\mu \over r\mu} K
- 2 r\left( 1 + {U\over r^2} - \sqrt{ 1 + 2{U\over r^2}}\right)\right].
\label{eq:es}
\end{equation}
From the explicit expressions for $K$ and $U$ it follows that $
E(r_H)<0$, and since the asymptotic behavior of the field
equations implies $F(r) \approx 1/r^2$ at infinity, we have that $
E\rightarrow +0$ at infinity. Therefore, the right hand side of
Eq.~(\ref{eq:es}) must be positive in some region. Moreover, there
must be a point where $ E=0$, i.e. $K = U$, and $E'>0$. At this
point, the right hand side of Eq.~(\ref{eq:es}) becomes
\begin{equation}
- r\mu F'^2 -2r \left ( \sqrt{ 1 + 2{ K \over r^2}} - 1 \right)
+ { K \over r\mu} (3\mu -1) >0 .
\end{equation}
Since the first and second terms of the last equation are
negative, we conclude that $ 3\mu >1$ at this point. This is the
same condition as in case $i)$.

$iii$) In the case of $SU(2)$ Yang--Mills--dilaton field with an
arbitrary (positive semidefinite) potential term $V(\phi)$ (which
is expected to arise in superstrings inspired models \cite{Hor2}),
the corresponding matter Lagrangian is given by \cite{eymd}
\begin{equation}
{\cal L}_{\rm MAT}=
{\cal L}_{\rm YMD}=
{\sqrt{-g}\over 4\pi}\left(
{1\over 2}\nabla_\mu \phi \nabla^\mu \phi
-{1\over 4 f^2} e^{2\gamma\phi}  {F_{\mu\,\nu}}^a\,{F^{\mu\,\nu}}_a
- V(\phi) \right) ,
\end{equation}
where $f$ is the gauge coupling constant, $\gamma$ is the
dimensionless dilatonic coupling constant, and  ${F_{\mu\nu}}^a$
is the SU(2) Yang--Mills field strength. The ansatz for the gauge
field configuration is the same as that given in case $i$), and
$\phi=\phi(r)$.

The corresponding field equations can be written in the generic
form (\ref{eq:ee}) with
$ K  =  K_1 +  K_2$, where
$ K _1= \mu\, \exp(2\gamma\phi) w'^2/f^2$,
$ K _2=\mu r^2 \phi'^2/2$ and
$U = r^2 V(\phi) + \exp(2\gamma\phi) (1-w^2)^2/(2f^2 r^2)$,
$\alpha = -2$, and $\beta=-2/(\mu r)$. Following the same
procedure, from the  matter field equations we find
\begin{equation}
 E' \equiv [ r^2 e^{-\delta} ( K  - U)]' =
r e^{-\delta}\left[-2\, K _2 - 4 r^2 V(\phi)
+ (3\mu -1)\, { K \over \mu} \right] \label{eq:eyd} .
\end{equation}
The behavior of the generic function $E$ is as in the previous
cases, and since the first and second terms of the right hand side
of Eq.~(\ref{eq:eyd}) are negative, we again find the condition
$3\mu > 1$ in order to obtain asymptotically flat solutions.

$iv$) For a $SU(2)$ Yang--Mills--Higgs field
 the matter Lagrangian is given by \cite{GreeneMO'N93}
\begin{equation}
{\cal L}_{\rm MAT}=
{\cal L}_{\rm YMH}=
-{\sqrt{-g}\over 4\pi} \,\left[
{1\over 4f^2} {F_{\mu\,\nu}}^a\,{F^{\mu\,\nu}}_a
+ (D_\mu \Phi)^\dagger\,(D^\mu \Phi) + V(\Phi)\right],
\label{eq:ymh}
\end{equation}
where $D_\mu $ is the usual gauge--covariant derivative, $\Phi$ is
a complex doublet Higgs field, and ${F^{\mu\,\nu}}_a$ is the
$SU(2)$ Yang--Mills field given above. The arbitrary potential
$V(\Phi)$ must be positive semidefinite. In this case, the ansatz
for the Yang--Mills field is the same as before, and for the Higgs
field we have
\begin{equation}
\Phi={1\over\sqrt{2}}\left(\matrix{0 \cr \varphi (r) \cr}\right).
\end{equation}
The field equations are equivalent to Eq.~(\ref{eq:ee}) with $ K
=  K_1 +  K_2$, where $ K _1=\mu r^2 \varphi'^2/2 $, and $ K _2=
\mu  w'^2/f^2$, $U = r^2 V(\varphi) + (1-w^2)^2/(2f^2 r^2) +
(1+w)^2\varphi^2/4$, $\alpha =- 2$, and $\beta=-2/(\mu r)$.
Finally, the matter field equations lead to
\begin{equation}
 E' \equiv [ r^2 e^{-\delta} ( K  - U)]' =
r e^{-\delta}\left[-2\, K _2
- 4 r^2 V(\varphi)
-{1\over 2} (1+w)^2 \varphi^2
+ (3\mu -1)\,{ K \over \mu} \right]  .
\end{equation}
As in the previous cases, the required behavior of the function
$E$ and the fact that the first three terms of the right hand side
of Eq.~(\ref{eq:ymh}) are negative lead to the condition $3\mu>1$
for the region of interest.

$v$) In the case of a non--Abelian Proca field the matter
Lagrangian is \cite{GreeneMO'N93}
\begin{equation}
{\cal L}_{\rm MAT} ={\cal L}_{\rm NAP} =
-{\sqrt{-g}\over 16\pi f^2}
 {F_{\mu\,\nu}}^a\,{F^{\mu\,\nu}}_a
-{\sqrt{-g}\,m^2\over 32\pi} A^a_\mu A_a^\mu\ ,
\label{eq:nap}
\end{equation}
where $m$ is the mass parameter and the ansatz for the potential
is as in the Yang--Mills case (\ref{eq:pot}). Again, the field
equations are given by Eq.~(\ref{eq:ee}) with $K = \mu w'^2 $, $U
= (1-w^2)^2/(2r^2) + m^2(1+w)^2$, $\alpha = -2/f^2$ and $\beta
=-2/(f^2\mu r)$. On the other hand, from the matter field
equations we obtain
\begin{equation}
 E' \equiv [ r^2 e^{-\delta} ( K  - U)]' =
r e^{-\delta}\left[ (3\mu -1) w'^2-{f^2m^2\over 2}(1+w)^2\right].
\end{equation}
As in the previous cases, the required behavior of the generic
function $E$ leads to the condition $3\mu > 1$.

In all cases presented here, there is a change in the behavior of
the generic function $ E$: It always starts at the horizon as a
negative and decreasing function and needs to increase towards its
asymptotic value. We have shown that this change always occurs
beyond the point characterized by $3\mu > 1$. Since $\mu =
1-2M(r)/r$, the change occurs at the point $r_0 > 3M(r_0)$. On the
other hand, $M(r)$ is an increasing function because from the
general field equation (\ref{eq:mu}) we have that $M'= -
(\widetilde{\kappa}/2) r^2 \Sigma^t_{\ t} = (\widetilde{\kappa}/2)
r^2 \rho_{\rm E} >0$, where $\rho_{\rm E}$ is the energy density
of the matter field which we suppose to be positive semidefinite
in accordance with the weak energy condition. Being an increasing
function, $M(r)$ reaches its minimum value on the horizon, where
$\mu(r_{\rm H})=0$ and $M(r_{\rm H}) = r_{\rm H}/2$. Consequently,
the turning point $r_0$ satisfies the inequality $r_0 > 3 M(r_0)
\geq 3 M(r_{\rm H}) = 3 r_{\rm H}/2$.

This result shows that the asymptotic behavior of the
matter fields present in the gravitational field of a
static spherically symmetric black hole can start only
after the value of the radial coordinate $r$ is sufficiently
large, and the lowest value determines the radius $r_{\rm hair}$
of the hairosphere. This proves the validity of the
no short hair conjecture for the triplet ansatz sector
of MAG in the presence of matter fields in which black
hole hair has been found.

\section{Discussion}

We have investigated the gravitational field of static spherically
symmetric black holes described by the effective Einstein--Proca
field of the triplet ansatz sector of MAG in the presence of
matter. It was shown that when matter is represented by an Abelian
Yang--Mills field, the no--hair theorems proven previously can be
applied. On the other hand, if the matter possesses a non--Abelian
structure or the corresponding Lagrangian contains an additional
potential term, the arguments employed to prove the validity of
no--hair theorems can be avoided due to the presence of an
additional term in the general matter field equations.

In particular, it was shown that black hole hair exists in the
system composed by the effective Einstein--Proca field of MAG and
a $SU(2)$ Yang--Mills, Skyrme, Yang--Mills--dilaton,
Yang--Mills--Higgs or non--Abelian Proca field. Moreover, we have
proved that in all these cases the no short hair conjecture is
valid, that is, hair exists only outside a sphere of radius
$r_{\rm hair} = 3r_{\rm H}/2$, where $r_{\rm H}$ is the horizon
radius.

These results could be used to further investigate the physical
significance of MAG. The no hair theorems proven in our previous
work \cite{eaah} show that the triplet ansatz sector of MAG in the
presence of a spherically symmetric black hole is nothing more but
Einstein's gravity. No new physics can be found in this sector
because the no hair theorems prohibit the existence of more
general solutions than the ones known in Einstein's gravity.
However, the triplet ansatz sector is probably one the most
simplest special cases of MAG. It could be that by slightly
relaxing the triplet ansatz (\ref{genEug}), one would obtain a
more general effective system which could be still equivalent to
Einstein's gravity coupled to a matter field. A first natural
candidate could be the effective Einstein--non--Abelian--Proca
field. In this case, as we have seen in this work, there exist
solutions with black hole hair. The hair could then be  directly
related to some specific parts of the post--Riemannian structures
of MAG. This research program, if realizable, could throw light on
the physical significance of torsion and nonmetricity.

\acknowledgments

We thank Friedrich W. Hehl for useful discussions and literature
hints. This research was supported by CONACyT Grants: 28339E,
32138E, by FOMES Grant: P/FOMES 98--35--15, by the joint
German--Mexican project CONACYT --- DFG: E130--655 --- 444 MEX 10,
and DGAPA--UNAM Grant 121298.



\begin{references}

\bibitem{Weyl} H. Weyl, {\em Naturwiss.} {\bf 19} (1931) 49.

\bibitem{quantum} E. S.\ Fradkin and A. A.\ Tseytlin, {\it Phys.\ Lett.}
{\bf B158} (1985) 316; C. G.\ Callan, D.\ Friedan, E. J.\ Martinec
and M. J.\ Perry, {\it Nucl.\ Phys.} {\bf 262} (1985) 593; D.\
Gross, {\it Phys.\ Rev.\ Lett.} {\bf 60} (1988) 1229; D.\ Gross
and P. F.\ Mende, {\it Nucl.\ Phys.} {\bf B303} (1988) 407 and {\it
Phys.\ Lett.} {\bf B197} (1987) 129.

\bibitem{PR} F. W. Hehl, J. D. McCrea, E. W.  Mielke, and Y. Ne'eman,
{\em Phys. Rep.} {\bf 258} (1995) 1. For computer algebra programs
see also: J. Socorro, A. Mac\'{\i}as, and F.W. Hehl, {\em Computer
Physics Communications} {\bf 115} (1998) 264.

\bibitem{hema99} F. W. Hehl and A. Mac\'{\i}as,
{\em Int. J. of Mod. Phys.} {\bf D8} (1999) 339.

\bibitem{obu} Yu. N. Obukhov, E. J. Vlachynsky, W. Esser,
and F. W. Hehl, {\em Phys. Rev.} {\bf D56} (1997) 7769.

\bibitem{tw} R. Tucker, C. Wang: Non--Riemannian gravitational
interactions. Talk given at {\em Mathematical Aspects of Theories
of Gravitation}, Warsaw, Poland (1996). Banach Centre Publications
Vol.\ {\bf 41} (Institute of Mathematics, Polish Academy of
Sciences, Warsaw 1997). Los Alamos E-Print Archive: gr-qc/9608055.

\bibitem{mms98} A. Mac\'{\i}as, E. W. Mielke, and J. Socorro,
{\em Class. Quantum Grav.} {\bf 15} (1998) 445.

\bibitem{ghlms} A. Garc\'{\i}a, F. W. Hehl, C. L\"ammerzahl,
A. Mac\'{\i}as, and J. Socorro, {\em Class. Quantum Grav.} {\bf
15} (1998) 1793. A. Garc\'{\i}a, A. Mac\'{\i}as, and J. Socorro,
{\em Class. Quantum Grav.} {\bf 16} (1999) 93. A. Mac\'{\i}as, C.
L\"ammerzahl, and A. Garc\'{\i}a, {\em J. Math. Phys.} {\bf 41}
(2000) 6369.

\bibitem{eaah} E. Ay\'on--Beato, A. Garc\'{\i}a, A. Mac\'{\i}as and
H. Quevedo, {\em Phys. Rev.} {\bf D61} (2000) 084017.

\bibitem{bek}  J. D. Bekenstein,
{\em Phys. Rev. Lett.} {\bf 28} (1972) 452;
{\em Phys. Rev.} {\bf D5} (1972) 1239.

\bibitem{GreeneMO'N93} B. R. Greene, S. D. Mathur, C. M. O'Neill,
{\em Phys. Rev.} {\bf D47} (1993) 2242.

\bibitem{Ayon00} E. Ay\'on--Beato,
{\em Phys. Rev.} {\bf D62} (2000) 104004.

\bibitem{ddh} D. N\'u\~nez, H. Quevedo and D. Sudarsky,
{\em Phys. Rev. Lett.} {\bf 76} (1996) 571.

\bibitem{Su} D. Sudarsky,
{\em Class. Quantum Grav.} {\bf 12} (1995) 579.

\bibitem{ym} J. A. Smoller, A. G. Wasserman and S. T. Yau,
{\em Comm. Math. Phys.}, {\bf 154} (1993) 377.

\bibitem{es} P. Bizon and T. Chamj,
{\em Phys. Lett.} {\bf B 297} (1992) 55;
M. Heusler, S. Droz, and N. Straumann,
{\em Phys. Lett.} {\bf B268} (1991) 371;
{\bf B271} (1991) 61; {\bf B258} (1992) 21.

\bibitem{Hor2} J. H. Horne and G. T. Horowitz,
{\em Phys. Rev.} {\bf D48} (1993) R5457.

\bibitem{eymd} G. Lavrelashvili and D. Maison,
{\em Nucl. Phys.} {\bf B 410} (1993) 407.

\end{references}
\end{document}